\documentclass[fleqn,usenatbib]{mnras}

\usepackage{newtxtext,newtxmath}

\usepackage[T1]{fontenc}
\usepackage{ae,aecompl}

\usepackage{graphicx}	
\usepackage{amsmath}	
\usepackage{pdflscape}
\usepackage{afterpage}
\usepackage{rotating}
\usepackage{xcolor}

\newcommand{\integral}{\textit{INTEGRAL}}

\newcommand{\nustar}{\textit{NuSTAR}}

\newcommand{\swift}{{\it Swift}}
\newcommand{\xmm}{{\it XMM-Newton}}

\newcommand{\red}{\textcolor{black}}
\newcommand{\ergs}{erg~cm$^{-2}$~s$^{-1}$}

\newcommand{\src}{MAXI~J1813}
\newcommand{\srcfull}{MAXI~J1813$-$095}

\newcommand{\second}{\textcolor{black}}

\title[Spectral Fitting of \srcfull]{A \nustar\ and \swift\ View of the Hard State of \srcfull}

\author[J. Jiang et al.]
{Jiachen Jiang$^{1,4}$\thanks{E-mail: jj447@cam.ac.uk}, Douglas J.K. Buisson$^{2}$, Thomas Dauser$^{3}$, Andrew C. Fabian$^{1}$, \newauthor Felix F{\"u}rst$^{5}$, Luigi C. Gallo$^{6}$, Fiona A. Harrison$^{7}$, Michael L. Parker$^{1}$, James F. Steiner$^{8}$,  \newauthor  John A. Tomsick$^{9}$, Santiago Ubach$^{8}$ and Dominic J. Walton$^{10,1}$\\
$^{1}$Institute of Astronomy, University of Cambridge, Madingley Road, Cambridge CB3 0HA, UK\\
$^{2}$Department of Astronomy, University of Southampton, Southampton SO17 1BJ, UK\\
$^{3}$Dr Karl Remeis-Observatory and Erlangen Centre for Astroparticle Physics, Sternwartstr. 7, D-96049 Bamberg, Germany\\
$^{4}$Department of Astronomy, Tsinghua University, 30 Shuangqing Road, Beijing 100084, China\\
$^{5}$European Space Agency (ESA), European Space Astronomy Centre (ESAC), E-28691 Villanueva de la Ca\~nada, Spain\\
$^{6}$Department of Astronomy and Physics, Saint Mary's University, 923 Robie Street, Halifax, NS B3H 3C3, Canada\\
$^{7}$Cahill Center for Astronomy and Astrophysics, California Institute of Technology, Pasadena, CA 91125, USA\\
$^{8}$Harvard-Smithsonian Center for Astrophysics, 60 Garden Street, Cambridge, MA 02138, USA\\
$^{9}$Space Sciences Laboratory, 7 Gauss Way, University of California, Berkeley, CA 94720-7450, USA\\
$^{10}$\red{Centre for Astrophysics Research, University of Hertfordshire, College Lane, Hatfield AL10 9AB, UK}}

\date{Accepted XXX. Received YYY; in original form ZZZ}

\pubyear{2021}

\begin{document}
\label{firstpage}
\pagerange{\pageref{firstpage}--\pageref{lastpage}}
\maketitle

\begin{abstract}
We present an analysis of the \nustar\ and \swift\ spectra of the black hole candidate \srcfull\ in a \red{failed-transition} outburst in 2018. The \nustar\ observations show evidence of reflected emission from the inner region of the accretion disc. By modelling the reflection component in the spectra, we find a disc inner radius of $R_{\rm in}<7$\,$r_{\rm g}$. This result suggests that either a slightly truncated disc or a non-truncated disc forms at a few per cent of the Eddington limit in \srcfull. Our best-fit reflection models indicate that the geometry of the innermost accretion remains consistent during the period of \nustar\ observations. The spectral variability of \srcfull\ from multi-epoch observations is dominated by the variable photon index of the Comptonisation emission.
\end{abstract}




\section{Introduction}

X-ray binaries are bright and variable X-ray emitters found in the Milky Way and nearby globular clusters \citep[e.g.][]{hertz83,jordan04}. Their X-rays are produced by the accretion process of materials from a donor star to a compact accretor, which is either a neutron star or a black hole (BH). Depending on the mass of the donor star, X-ray binaries are classified as low-mass and high-mass X-ray binaries. The donor star of a low-mass X-ray binary is usually less massive than the accretor, e.g. dwarf stars or main-sequence stars \citep[e.g.][]{liu07}. In comparison, the donor of a high-mass X-ray binary is more massive, e.g. blue giants \citep[e.g.][]{liu06}. 

Some X-ray binaries show transient events featured by a boost of accretion rate along with a rapid increase of X-ray luminosity. The `soft' state and the `hard' state are the two most distinct spectral states in \second{such} outburst: the X-ray emission of an X-ray binary is dominated by the thermal emission of the accretion disc in the soft state and the non-thermal power-law emission from the corona in the hard state \citep[e.g.][]{oda71,cui98,rao00,yu09,shidatsu11,reis13}. Such a change of state often shows the so-called `q'-shaped pattern in the X-ray hardness-intensity diagram \citep[e.g.][]{fender04, homan05,dunn10}. Other intermediate states have also been identified during the transition phases between the canonical `soft' and `hard' states \citep[e.g.][]{done07}. A few sources among known X-ray binaries never reached the soft state during an outburst \citep[e.g.][]{brocksopp04, sturner05, jiang20}, which is often referred to as a \red{failed-transition} outburst.

\red{In the `no-hair theorem', an astrophysical
BH is described by just its mass and spin \citep[for a review, see][the third parameter electrical charge being effectively neutral in an astrophysical setting]{chrusciel12}. Understanding BH spin is important in different aspects. For example, the spins of the BHs have been proposed to be the energy source behind energetic jets launched from BHs. This assumption was supported  by both theories and observations \citep[e.g.][]{blanford77,tchekhovskoy11,mcclintock14}. Two techniques have been commonly used to measure BH spins, the X-ray continuum-fitting method \citep{zhang97} and
 the relativistic disc reflection spectroscopy \citep{fabian89}. They both rely on the monotonic relationship between
BH spin and the radius of the innermost stable circular orbit (ISCO) for orbiting particles around a BH. It is, therefore, essential to study whether the disc extends to the ISCO or is truncated during a certain observation, which determines the validity of the two methods for BH spin measurements.} 

\red{\citet{esin97} suggested that the inner accretion disc of a BH X-ray binary might be indeed truncated in a low luminosity state. Instead, an advection-dominated accretion flow onto the BH appears in the low-flux/hard state \citep[e.g.][]{quataert99}. But the degree to which the discs of BH X-ray binaries are truncated in the hard state at a modest luminosity, e.g. 1\%-10\% of $L_{\rm Edd}$, is still a controversial question. Recent work on this topic focused on the modelling of the disc reflection spectra in the X-ray band: some found discs are either close to ISCO or slightly truncated \citep[e.g. in GX~339$-$4,][]{garcia15,furst15}; while others found that the discs are significantly truncated  \citep[e.g. 300\,$r_{\rm g}$ for GX~339$-$4,][]{plant15}. Intermediate values of $R_{\rm in}$, e.g. a few tens of $r_{\rm g}$, were also obtained \citep[e.g. Cyg~X--1,][]{basak17}. It is important to note that, if the thin disc is truncated at 100-300\,$r_{\rm g}$, the coronal region, which illuminates the disc, has to extend to a similar size to produce the observed amount of reflection. Such a large size would bring challenges in explaining the steep disc emissivity profiles \citep{fabian12} and the observed reverberation lags in the hard state \citep{demarco13,kara19}. The diametrically opposed conclusions might be due to calibration issues in timing-mode data from CCDs, e.g. on \xmm\, and the usage of different models \citep{kolehmainen14,garcia15}.} 


\begin{figure}
    \centering
    \includegraphics[width=9cm]{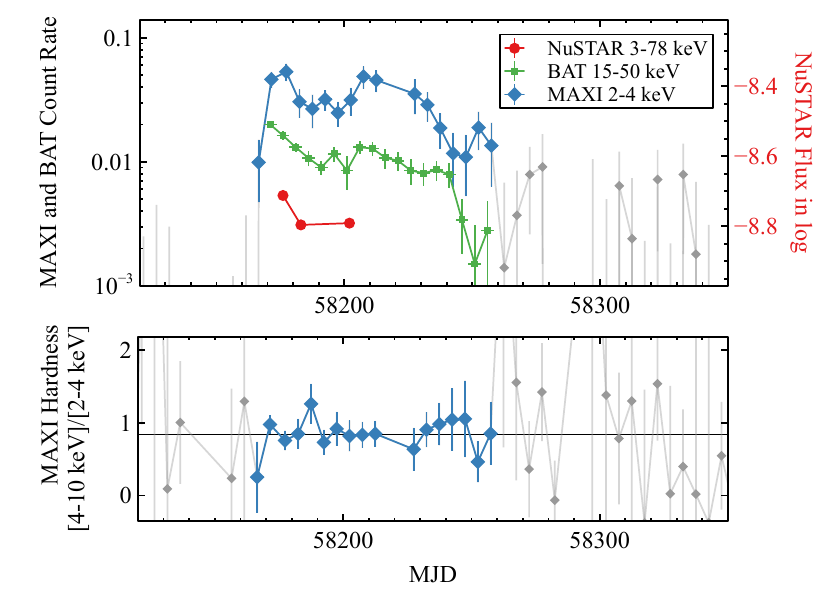}
        \label{pic_lc}
    \caption{Top: \textit{MAXI} (blue \red{and grey} diamonds), \swift-BAT (green squares) and \nustar\ lightcurves (red circles) of \src\ in outburst in 2018. The error bars of the \nustar\ lightcurves are smaller than the size of the circles. Bottom: \textit{MAXI} hardness ratio curve of \src.}
\end{figure}

\red{In this work, we study the geometry of the innermost accretion region of \srcfull\ in the hard state by modelling its broad band X-ray spectra.} \srcfull\ (\src\ hereafter) is an X-ray binary discovered by \textit{MAXI} that had a \red{failed-transition} X-ray outburst in 2018 \citep{kawase18}. The nature of the compact object in \src\ has not been dynamically confirmed. However, the X-ray band of \src\ is dominated by non-thermal power-law emission as in the hard state of typical BH transients \citep{furst18,armas19}.   


Figure\,\ref{pic_lc} shows the X-ray lightcurves of \src\ in 2018. \src\ quickly reached the peak X-ray flux within the first 10 days. Then the outburst lasted for around 90 days before returning to the quiescent state. The lower panel of Figure\,\ref{pic_lc} shows the \textit{MAXI} hardness ratio curve for the same period \citep{matsuoka09}, which remains consistent during the whole outburst\footnote{Interested readers can find the \swift\ hardness-intensity diagram of this outburst in \citet{jana21}.}. Detailed spectral analysis of \integral, \textit{NICER} and \xmm\ data suggests that \src\ remains in the hard state during the outburst \citep{furst18, armas19, jana21}.

In this work, we analyse three sets of \nustar\ and \swift\ observations of \src\ in outburst. We focus on the modelling of their X-ray spectra. In particular, we try to probe the geometry of the innermost accretion region of \src\ in the hard state, e.g. whether the disc is truncated at a large radius \citep[e.g.][]{plant15} or close to the BH \citep[e.g.][]{garcia15}. 


In Section\,\ref{reduc}, we introduce our data reduction processes. In Section\,\ref{stack}, we present a detailed analysis of the stacked \nustar\ and \swift\ spectra of \src\ considering various reflection models. In Section\,\ref{each}, we study the multi-epoch variability of \src\ based on the best-fit reflection model. In Section\,\ref{conclude}, we conclude our results.

\section{Data Reduction} \label{reduc}

\subsection{\nustar}

We reduced the \nustar\ data using the \nustar\ Data Analysis Software (NuSTARDAS) package and calibration data of V20200510. The energy spectra of \src\ were extracted for both the FPMA and FPMB detectors from a 100$''$ radius circle centered on the source, while the background spectra were extracted from source-free polygon regions on the same detector chip. We consider the 3--78\,keV band of the two FPM spectra. 

\subsection{\swift}

{The X-ray Telescope (XRT) data from the \swift\ observation were reduced using XRTPIPELINE version 0.13.4. The calibration file version is V20200726. The observation was operated in the Window Timing mode. Source spectra were extracted from a radius of 20 pixels\footnote{Each pixel is approximately $2.36^{\prime\prime}$.}. The background region was chosen to be an annular region with an inner radius of 50 pixels and an outer radius of 70 pixels. The 0.5--10\,keV band of the XRT data is considered in this work.}

All the spectra are grouped to have a minimum signal-to-noise of 6 per bin and oversample by a factor of 3. We use XSPEC V12.11.1 \citep{arnaud96} for spectral analysis, and $\chi^{2}$ is used for the goodness-fit test in this work. \second{The uncertainties are given at the 90\% confidence level unless specifically mentioned otherwise.}

\begin{table*}
    \centering
    \caption{A list of observations of \src\ taken in 2018. \red{The last three columns show the observed flux of \src\ in the 1--3, 3--10 and 10--78\,keV bands in units of $10^{-10}$\,\ergs. The 3--10 and 10--78\,keV flux of \src\ are the mean values of FPMA and FPMB measurements. The 3--10\,keV flux measured by XRT is shown in the brackets. The 1--3\,keV flux is measured by XRT.}}
    \begin{tabular}{ccccccccc}
    \hline\hline
    \nustar & Date & Exposure & \swift & Date & Exposure & \red{$F_{\rm 1-3 keV}$} & \red{$F_{\rm 3-10 keV}$} & \red{$F_{\rm 10-78 keV}$}\\
    & & & & & & XRT & FPM (XRT) & FPM \\
    & & (ks) & & & (ks) & {($10^{-10}$\,\ergs)} & {($10^{-10}$\,\ergs)} &{($10^{-10}$\,\ergs)}\\
    \hline
    80402303002 & 02-28 & 20.5 & 00088654001 & 02-27 & 1.9  & $1.55\pm0.02$ & $4.21\pm0.02$ ($4.31\pm0.05$) & $15.20\pm0.02$\\
    80402303004 & 03-06 & 20.4 & 00088654002 & 03-06 & 1.8 & $1.28\pm0.02$ & $3.50\pm0.02$ ($3.56\pm0.04$) & $12.51\pm0.02$\\
    80402303006 & 03-25 & 23.2 & 00088654004 & 03-25 & 2.0 & $1.49\pm0.02$ & $3.84\pm0.02$ ($3.87\pm0.05$) & $12.32\pm0.02$\\
    \hline\hline
    \end{tabular}
    \label{tab_obs}
\end{table*}

\section{Stacked Spectral Analysis} \label{stack}

\begin{figure}
    \centering
    \includegraphics[width=8cm]{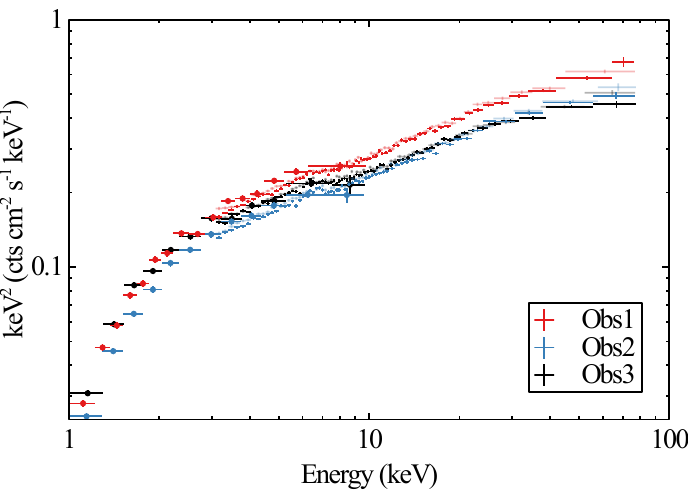}
    \caption{\red{X-ray spectra of \src\ (Red: Obs1; blue: Obs2; black: Obs3). The spectra are corrected for the effective area of the instruments but not unfolded from the instrumental response.  Data from XRT, FPMA and FPMB are plotted, but shown in the same colour to ease comparison between epochs. Crosses: FPMA; fainter crosses: FPMB; circles: XRT.}}
    \label{pic_folded}
\end{figure}

\red{The hardness ratio of \src\ remained at a similar level in the period of three \nustar\ observations: the flux ratio between 3--10\,keV and 1--3\,keV bands is around 2.6--2.7 measured by \nustar\ \second{(see Table\,\ref{tab_obs})}. A quick view of the spectra is given in Fig.\,\ref{pic_folded}.  The spectra of the three epochs share a similar spectral shape.} We therefore start our analysis with the averaged spectra of \src\ averaged from all three epochs. \red{The ADDSPEC tool is used to stack spectra.} 

\begin{figure}
    \centering
    \includegraphics[width=9cm]{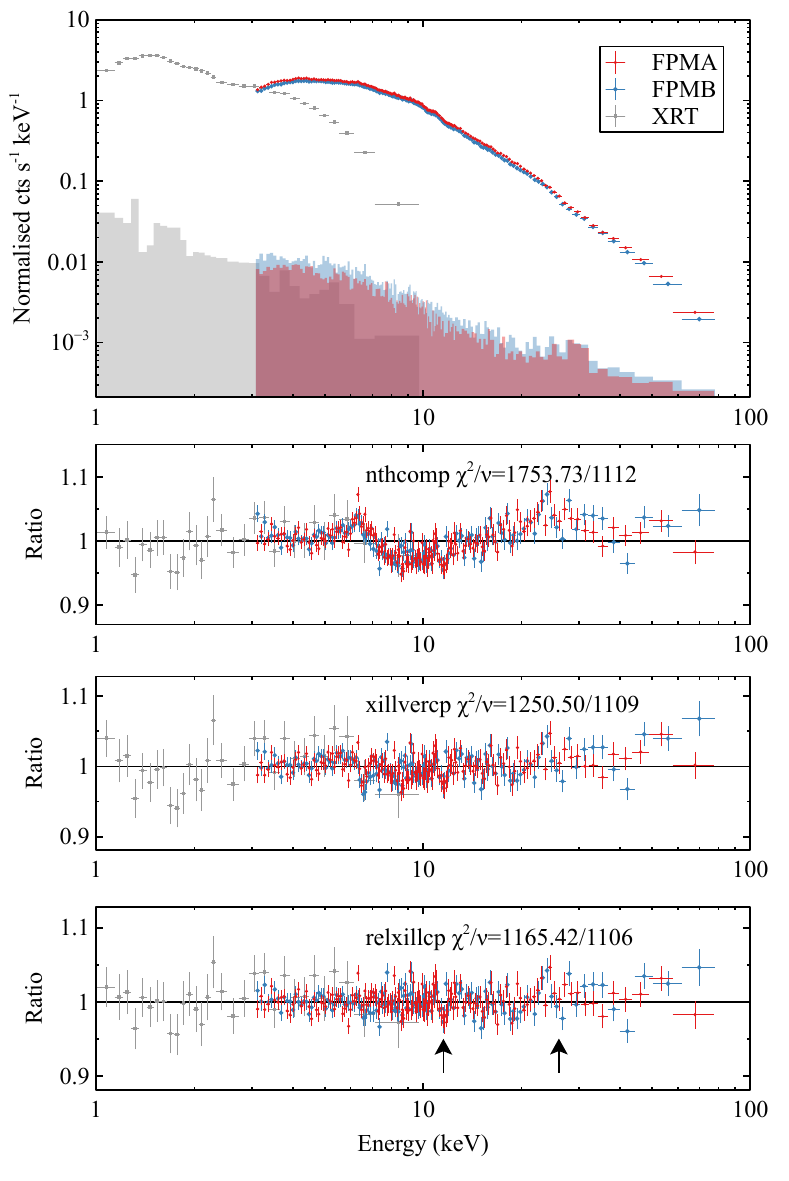}
    \caption{Top: \nustar\ (red: FPMA; blue: FPMB) and \swift\ (grey: XRT) source (crosses) and background spectra (shaded regions) of \src. Bottom three panels: data/model ratio plots using different models. FPM spectra show evidence of broad emission line around 6.4\,keV and a hump feature above 10\,keV, suggesting existence of a reflection component. A distant, ionised reflector (\texttt{xillvercp}) fails to fit the spectra with significant residuals around the iron emission band as shown in the third panel. A relativistic disc reflection model (\texttt{relxillcp}) is therefore used to improve the fit. The residuals at 12 and 28\,keV marked by the black arrows are due to instrumental features \citep{madsen15}.  }
    \label{pic_data}
\end{figure}

To start with, we fit the spectra with an absorbed Comptonisation model. The \texttt{nthcomp} model is used to calculate the Comptonisation spectrum \citep{zycki99}, and the \texttt{tbnew} model is used to account for Galactic absorption \citep{wilms06}. The data/model ratio plot is shown in Figure\,\ref{pic_data}. Evidence of a broad emission line around 6.4\,keV and a hump feature above 10\,keV is found in the \nustar\ spectra, suggesting the existence of a reflection component in the data. An additional distant, ionised reflector \texttt{xillvercp} \citep{garcia10} fails to fit the broad emission line with significant residuals in the iron emission band (see the third panel). This suggests that the reflection component originates in the inner region of the accretion disc where relativistic corrections are required \citep[e.g.][]{fabian89}. 

In this section, we introduce three models for the disc reflection component in the spectra of \src, one with a power-law disc emissivity profile, one for a disc illuminated by an isotropic point-like corona, i.e. \red{in the} `lamppost' geometry and \second{a high-density disc reflection model}.  

\subsection{\texttt{relxillcp}}

\begin{figure}
    \centering
    \includegraphics[width=7cm]{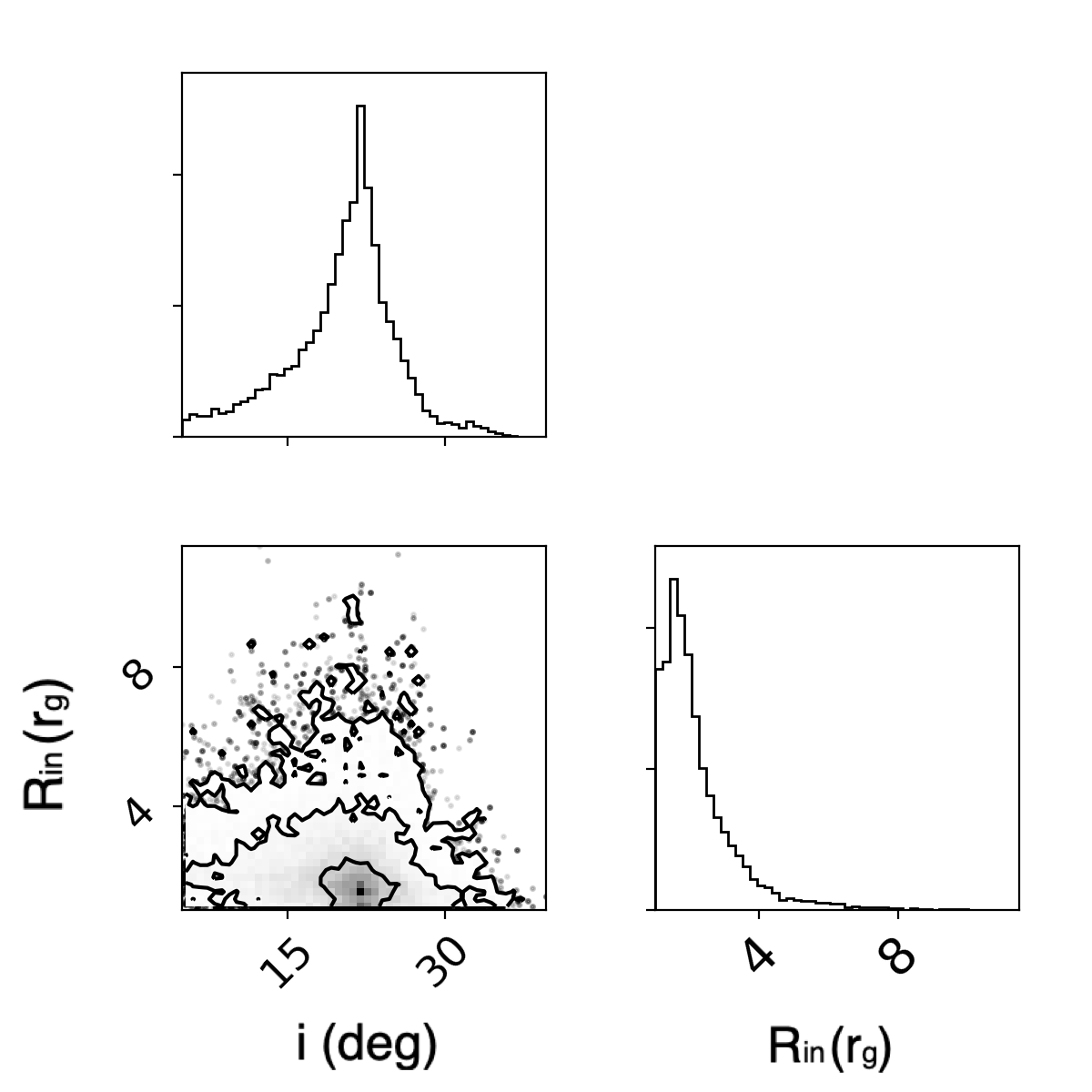}
    \caption{Output distributions of MCMC analysis based on the \texttt{relxillcp} model. Contours correspond to 1, 2 and 3$\sigma$ uncertainty ranges.}
    \label{pic_cp_mc}
\end{figure}

\begin{table}
    \centering
    \caption{Best-fit parameters for the averaged \nustar\ and \swift\ spectra of \src. Note that the values of $f_{\rm refl}$ of \texttt{relxillcp} and \texttt{relxilllpcp} are not comparable as different geometries of the corona are assumed. \second{The quoted errors are at the 90\% confidence level.}}
    \label{tab_cp}
    \begin{tabular}{cccc}
    \hline\hline
    Parameters  & \texttt{relxillcp} & \texttt{relxilllpcp} & \texttt{reflionx} \\
    \hline
    $N_{\rm H}$ ($10^{21}$\,cm$^{-2}$) & $8.09\pm0.02$ & $8.091^{+0.019}_{-0.016}$ & $9.50\pm0.04$\\
    \hline
    q & $2.5^{+0.7}_{-0.3}$ &- & - \\
    h ($r_{\rm g}$) & - & $11^{+6}_{-4}$ & $14^{+7}_{-5}$ \\
    $i$ (deg) & $22^{+8}_{-9}$ & $23^{+6}_{-11}$ & $20^{+7}_{-10}$\\
    $R_{\rm in}$ ($r_{\rm g}$) & <5 & <7 & <7 \\
    $\Gamma$ & $1.644\pm0.006$ & $1.644\pm0.007$ & $1.682^{+0.005}_{-0.007}$ \\
    $kT_{\rm e}$ (keV) & >180 & >190 & > 170 \\
    $Z_{\rm Fe}$ ($Z_{\odot}$) & $1.6^{+0.4}_{-0.6}$  & $1.7^{+0.4}_{-0.5}$ & $1.1\pm0.3$\\
    $n_{\rm e}$ (cm$^{-3}$) & $10^{15}$ & $10^{15}$ & $10^{19}$\\
    $f_{\rm refl}$ & $0.08\pm0.02$ & $0.19^{+0.02}_{-0.04}$ & -\\
    \red{$\log(\xi$/erg\,cm\,s$^{-1})$} & $3.10^{+0.09}_{-0.06}$ & $3.10^{+0.07}_{-0.06}$ & $2.13^{+0.12}_{-0.09}$ \\ 
    Norm ($10^{-3}$) & $4.31\pm0.03$ & $5.8^{+0.3}_{-0.4}$ & $2.4^{+2.3}_{-0.4}\times10^3$\\
    $\rm Norm_{pl}$ & -& - & $0.11\pm0.03$ \\
    \hline
    $C_{\rm FPMB}$ & $1.025\pm0.003$ & $1.025\pm0.003$ & $1.024\pm0.003$ \\
    $C_{\rm XRT}$ & $1.03\pm0.02$ & $1.03\pm0.03$ & $1.03\pm0.03$ \\
    \hline
    $\chi^{2}/\nu$ & 1165.42/1106 & 1166.35/1106 & 1164.27/1106 \\
    \hline\hline
    \end{tabular}
\end{table}

We first apply the \texttt{relxillcp} model to the spectra of \src\ \citep{dauser13, garcia10}. This model calculates relativistic disc reflection spectra by given seed photon spectra in the shape of \texttt{nthcomp}. We allow the reflection fraction ($f_{\rm refl}$) in the model to be a positive, free parameter in our spectral fitting. So, the model includes both the disc reflection component and the coronal Comptonisation component. Other free parameters include the inner disc radius\footnote{The BH spin parameter is fixed at 0.998.} ($R_{\rm in}$), the inclination angle of the disc ($i$), the emissivity index (q), the iron abundance of the disc ($Z_{\rm Fe}$) and the ionisation of the disc ($\xi$). A constant density of $n_{\rm e}=10^{15}$\,cm$^{-3}$ is assumed for the disc in this model. A \texttt{constant} model is used to account for cross-calibration uncertainties between instruments. The full model is \texttt{constant * tbnew * relxillcp} in XSPEC notation.

\red{In this \texttt{relxillcp} model, we calculate the reflection spectrum of the disc taking all the relativistic effects into account, and the } fit is significantly improved with $\Delta\chi^{2}=85$ and three more free parameters \red{compared to the fit using the distant reflection model \texttt{xillvercp}}. Best-fit parameters are given in the first column of Table\,\ref{tab_cp}, and corresponding data/model ratio plots are shown in the bottom panel of Figure\,\ref{pic_data}. \red{The values of the \texttt{constant} models are within the expectations based on the cross-calibration work in \citep{madsen15}.}

We add an additional \texttt{diskbb} component to account for any possible disc thermal emission. The fit is not significantly improved $\Delta\chi^{2}<2$ and two more free parameters. We, therefore, conclude that no significant thermal emission from the disc is found in our data (see Appendix\,\ref{bb_sec} for more discussion). \red{We note that a disc thermal component was identified by \citet{jana21} using the same \nustar\ and \swift\ observations. The requirement for an additional disc thermal component is based on a simple phenomenological model where Fe~K emission is fit by a relativistic disc line model. The contribution of disc reflection in the soft and hard X-ray band is ignored.} 

The best-fit \texttt{relxillcp} model suggests a thin disc with an inner radius of $R_{\rm in}<5$\,$r_{\rm g}$ forms around the BH during our observations. The small value of $R_{\rm in}$ indicates either a slightly truncated disc or a disc that extends to the innermost stable circular orbit exists in the hard state of \src. Besides, the model implies an almost face-on viewing angle of 22$^{\circ}$ for the disc. 

We further estimate measurement uncertainties using the MCMC algorithm. The XSPEC/EMCEE code based on \citep{foreman12} of the Goodman-Weare affine invariant MCMC ensemble sampler \citep{goodman10} is used for this purpose. We use 200 walkers with a length of 100000, burning the first 10000. A convergence test has been conducted and the Gelman-Rubin scale-reduction factor $R<1.3$ for every parameter. No obvious degeneracy is found. The contour plots of $R_{\rm in}$ and $i$ are shown in Figure\,\ref{pic_cp_mc}. The uncertainties given by MCMC analysis are consistent with the values obtained by the ERROR command in XSPEC. For instance, the 3-$\sigma$ upper limit of $R_{\rm in}$ is 8\,$r_{\rm g}$ when using the \texttt{relxillcp} model.

\subsection{\texttt{relxilllpcp}}

So far, we have obtained a good fit using the \texttt{relxillcp} model. A power-law emissivity profile of ($F\propto r^{-q}$)is used in this model, and no particular geometry is assumed for the coronal region\footnote{Disc emissivity profiles were calculated for various coronal geometries, e.g. sphere and jet-like \citep{gonzalez17}. A power law or a broken power law is found to be a good approximation for their emissivity profiles \citep{wilkins12,gonzalez17}.}.

In this section, we consider the lamppost geometry for the innermost region of the disc \citep{martocchia96}. In particular, we investigate whether the choice of the lamppost geometry affects our measurements of the geometry of the innermost accretion region in \src. The \texttt{relxilllpcp} model is used \citep{dauser16} for this purpose. Instead of a power-law emissivity profile, \texttt{relxilllpcp} calculates emissivity profiles depending on $h$, the height of the corona above the BH on its rotational axis.

\begin{figure}
    \centering
    \includegraphics[width=8.5cm]{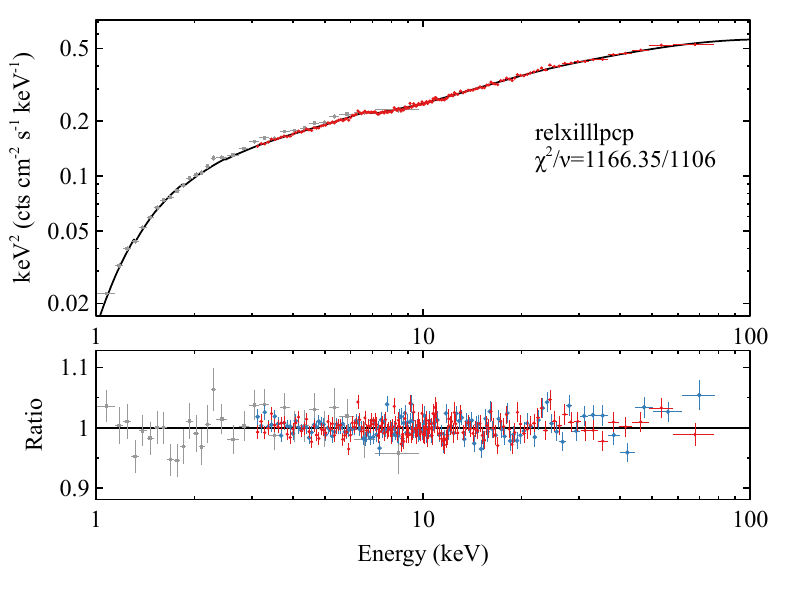}
    \caption{Top: the best-fit \texttt{relxilllpcp} model (black line) and unfolded FPMA (red) and XRT (grey) spectra of \src. Bottom: corresponding data/model ratio plots.}
    \label{pic_lpcp_fit}
\end{figure}

By applying the \texttt{relxilllpcp} model to the spectra, we also find a good fit with $\chi^{2}/\nu=1166.35/1106$. The goodness of the fit is consistent with that of \texttt{relxillcp}. The best-fit model and corresponding data/ratio plots are shown in Figure\,\ref{pic_lpcp_fit}.  MCMC analysis is also used to estimate measurement uncertainties. 

\begin{figure}
    \centering
    \includegraphics[width=8.5cm]{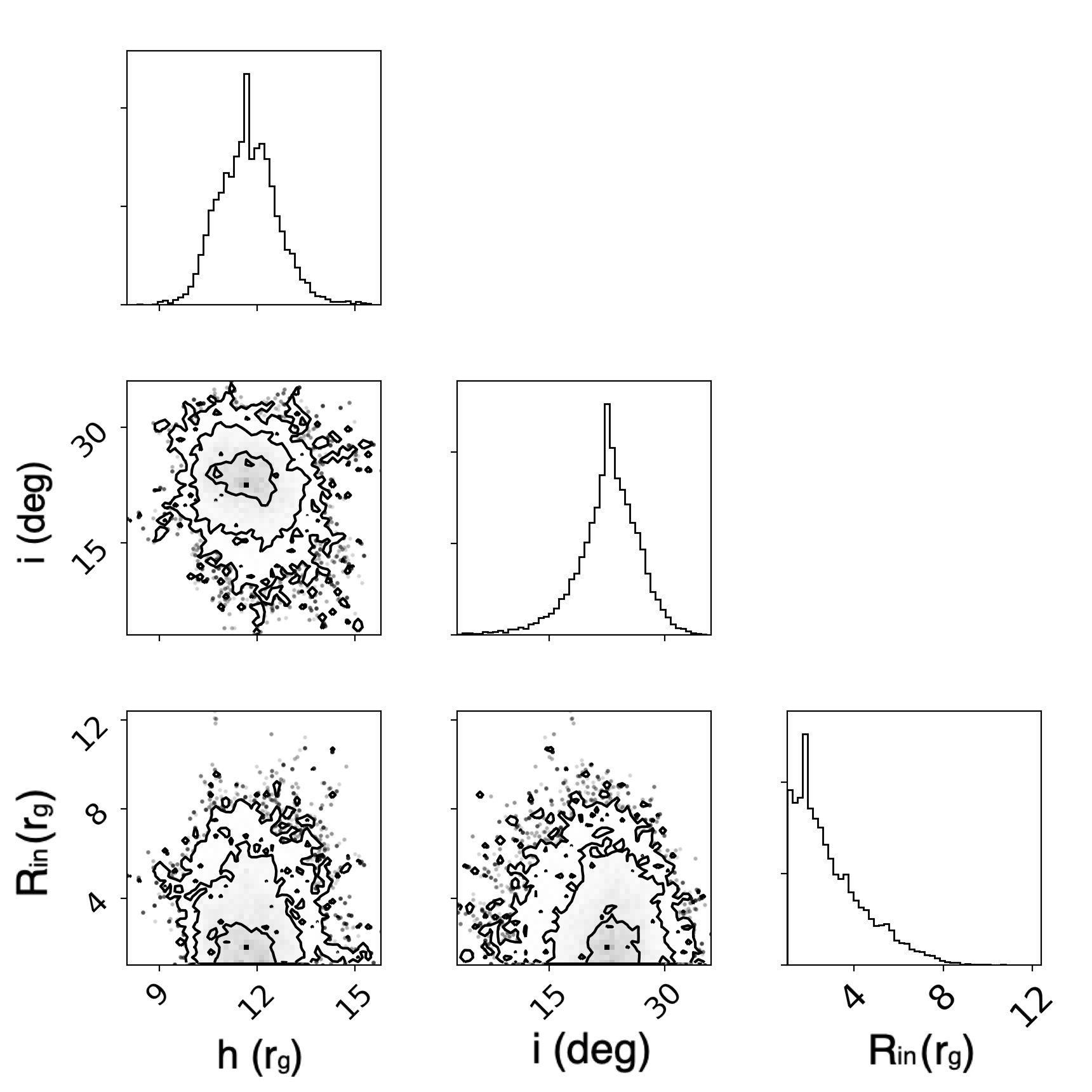}
    \caption{Same as Figure \ref{pic_cp_mc} but based on the \texttt{relxilllpcp} model.}
    \label{pic_lpcp_mc}
\end{figure}

When the \texttt{relxilllpcp} model is used, the upper limit of $R_{\rm in}$ increases slightly: the 2-$\sigma$ upper limit of $R_{\rm in}$ is 6\,$r_{\rm g}$ (see Fig.\,\ref{pic_lpcp_mc}). In comparison, the 2-$\sigma$ upper limit given by the \texttt{relxillcp} model is 4\,$r_{\rm g}$. However, their measurements of $R_{\rm in}$ are \red{similar}. The inferred inclination angles from two models are also consistent. Our lamppost model also suggests a coronal region that extends to $h=11^{+3}_{-4}$\,$r_{\rm g}$.

\subsection{\texttt{reflionx}} \label{ne}

Recently, it has been found that a high disc electron density of over $n_{\rm e}=10^{19}$\,cm$^{-3}$ is required to explain the reflection spectra of BH transients in various states \citep{tomsick18,jiang19,jiang20,connors21}. This model provides a possible solution to inferred supersolar iron abundances in previous reflection models \citep{tomsick18,jiang18,jiang19b}. 

We test a high density disc reflection model with the \texttt{reflionx} model \citep{ross07}. The \texttt{reflionx} is calculated with \texttt{nthcomp}-shaped illuminating spectra \citep{jiang20b}. The \texttt{relconvlp} model is applied to the \texttt{reflionx} model \red{to account} for relativistic corrections. The full model is \texttt{constant * tbnew * (relconvlp * reflionx + nthcomp)} in XSPEC notation. In comparison with the \texttt{relxilllpcp}, this model has one additional parameter, the electron density of the disc surface ($n_{\rm e}$). 

\begin{figure}
    \centering
    \includegraphics[width=8.5cm]{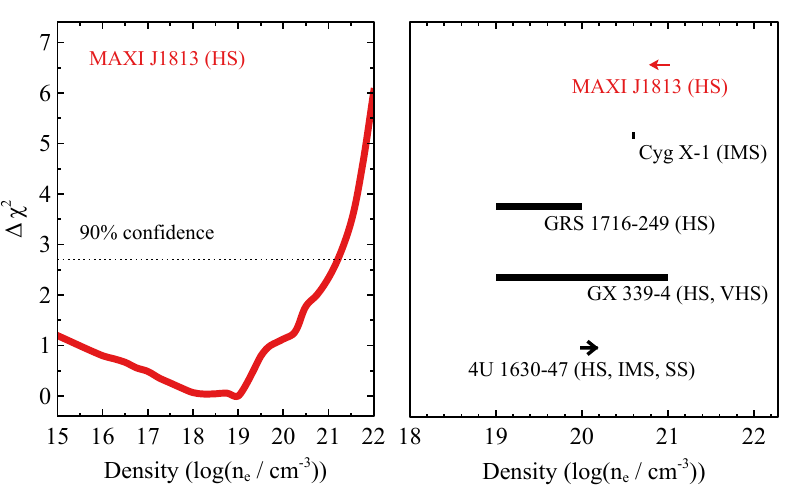}
    \caption{Left: $\chi^{2}$ vs. $n_{\rm e}$ obtained by fitting the spectra of \src\ with \texttt{reflionx}. Only an upper limit of $n_{\rm e}$ ($<10^{21}$\,cm$^{-3}$) is found for the disc in \src. Right: $n_{\rm e}$ of \src\ in comparison with the measurements of other BH transients \citep{tomsick18,jiang19,jiang20,connors21}. HS: the hard state; IMS: the intermediate state; VHS: the very high state; SS: the soft state.}
    \label{pic_ne}
\end{figure}

By applying the \texttt{reflionx} model to the data, we only find an upper limit for the $n_{\rm e}$ parameter of $10^{21}$\,cm$^{-3}$. The $\chi^{2}$ distribution against $n_{\rm e}$ is shown in Figure\,\ref{pic_ne}. Tentative evidence of $n_{\rm e}=10^{19}$\,cm$^{-3}$ is suggested by our fit. But $\Delta\chi^{2}$ is lower than 1.2 in the range of $10^{15}-10^{20}$\,cm$^{-3}$. The poorly constrained density parameter is likely due to the lack of evidence of blackbody-like emission in the \swift\ data of \src. As shown in Figure\,\ref{pic_data}, the XRT spectrum is consistent with an absorbed power law below 3\,keV. At a high density, the disc reflection spectrum shows a blackbody-like emission due to stronger free-free absorption. 

We show the upper limit of $n_{\rm e}$ of \src\ in comparison with other BH transients in the right panel of Figure\,\ref{pic_ne}. GX~339$-$4 and GRS~1716$-$249 show a variable density parameter during their outburst \citep{jiang19,jiang20}. A lower limit of $n_{\rm e}=10^{20}$\,cm$^{-3}$ is found in all spectral states of 4U~1630$-$47 \citep{connors21}. 

\begin{figure}
    \centering
    \includegraphics[width=8.5cm]{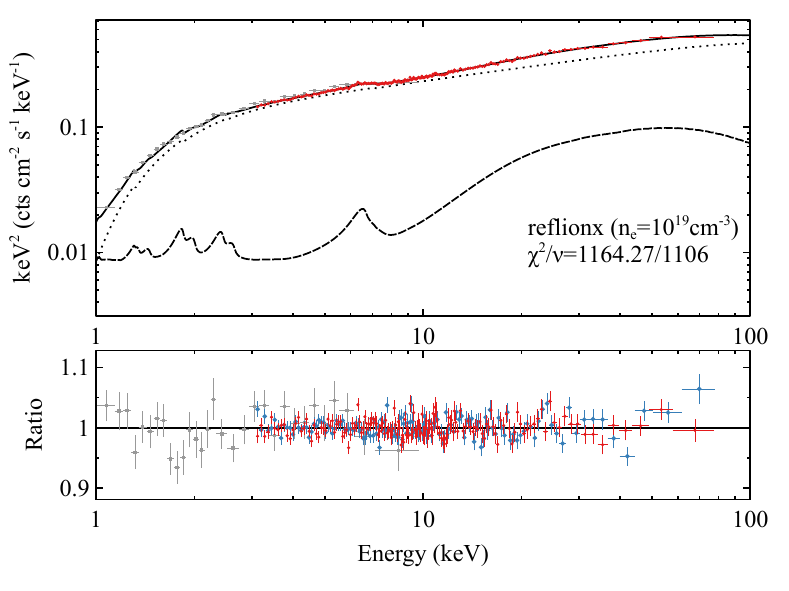}
    \caption{Top: the best-fit model with $n_{\rm e}=10^{19}$\,cm$^{-3}$ (solid black line) and unfolded FPMA (red) and XRT (grey) spectra of \src. Dotted line: the Comptonisation component; dash-dotted line: the disc reflection component. Bottom: corresponding data/model ratio plots.}
    \label{pic_ne_fit}
\end{figure}

We further investigate whether a high density model would affect our measurements of $R_{\rm in}$ and $i$. We fix $n_{\rm e}=10^{19}$\,cm$^{-3}$, where the minimum $\chi^{2}$ is found. Then we fit the spectra with all the other parameters free to vary. We obtain a similar upper limit for  $R_{\rm in}$ (<7$r_{\rm g}$) and a similar disc inclination angle ($i\approx20^{\circ}$).  The best-fit parameters are shown in the last column of Table\,\ref{tab_cp}, and the best-fit model is shown in Figure\,\ref{pic_ne_fit}. Most of the parameters are consistent with the values obtained by the \texttt{relconvlp} model except for the \second{ionisation parameter of the disc and the} line-of-sight column density. \second{The ionisation parameter ($\log(\xi)$) of this high-density model is lower than the inferred values of best-fit low-density models.} A slightly higher column density of $N_{\rm H}=9.50\pm0.04\times10^{21}$\,cm$^{-2}$ is found\footnote{The line-of-sight Galactic column density of H~\textsc{I} and H$_{2}$ combined is estimated to be $5.4\times10^{21}$\,cm$^{-2}$ \citep{willingale13}, which is lower than the inferred values from our observations. Similar results were found in \xmm\ and \integral\ spectral analysis \citep{furst18,armas19}.}. This is due to the blackbody-like emission in the high density disc model, where free-free absorption increases the temperature of the disc surface \citep{ross07,garcia16}. A slightly higher $N_{\rm H}$ is therefore required when $n_{\rm e}=10^{19}$\,cm$^{-3}$ is used in the model.

\section{Individual Observations} \label{each}

\begin{figure}
    \centering
    \includegraphics[width=8.5cm]{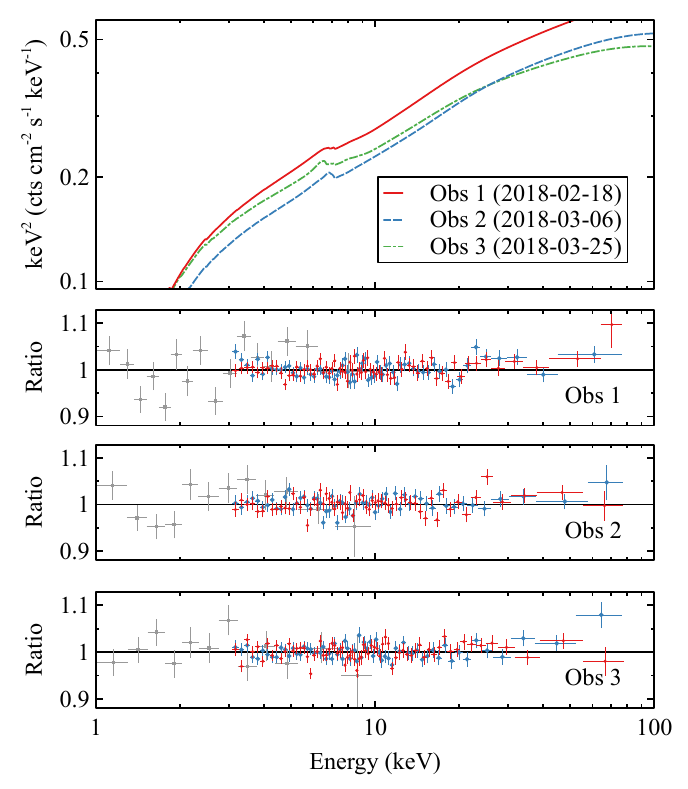}
    \caption{Top: best-fit \texttt{relxilllpcp} models for each individual observation. Bottom: corresponding data/model ratio plots for each epoch. $F_{\rm 3-78keV}$ is absorption-corrected X-ray flux in the \nustar\ band (3--78\,keV).}
    \label{pic_3obs}
\end{figure}

\begin{table}
    \centering
    \caption{Best-fit parameters for each observation of \src. \red{$F_{\rm 1-78keV}$ is the unabsorbed flux of \src\ in units of $10^{-9}$\,\ergs.} \second{The quoted errors are at the 90\% confidence level.}}
    \label{tab_3obs}
    \begin{tabular}{cccc}
    \hline\hline
    Parameters  & Obs1 & Obs2 & Obs3\\
    \hline
    $N_{\rm H}$ ($10^{21}$\,cm$^{-2}$) & $8.21\pm0.02$ & $8.30^{+0.02}_{-0.03}$ & $7.92\pm0.05$ \\
    \hline
    h ($r_{\rm g}$) & $14^{+9}_{-10}$ & $18^{+6}_{-12}$ & $18^{+22}_{-14}$ \\
    $i$ (deg) & - & $23\pm10$  & - \\
    $R_{\rm in}$ ($r_{\rm g}$) & <9 & <14 & <13 \\
    $\Gamma$ & $1.620^{+0.015}_{-0.012}$ & $1.651\pm0.007$ & $1.681^{+0.011}_{-0.010}$ \\
    $kT_{\rm e}$ (keV) & >200 & >180 & >150\\
    $Z_{\rm Fe}$ ($Z_{\odot}$) & - & $1.5\pm0.5$ & - \\
    $f_{\rm refl}$ & $0.18\pm0.03$ & $0.18^{+0.02}_{-0.04}$ & $0.18^{+0.04}_{-0.03}$ \\
    \red{$\log(\xi$/erg\,cm\,s$^{-1})$} & $3.18^{+0.10}_{-0.12}$ & $2.8\pm0.2$ & $3.11^{+0.11}_{-0.20}$ \\ 
    Norm ($10^{-3}$) & $6.5^{+0.6}_{-1.4}$  & $4.7^{+0.2}_{-0.4}$ & $4.3^{+0.6}_{-0.4}$\\
    \hline
    $C_{\rm FPMB}$ & $1.020\pm0.004$ & $1.017\pm0.004$ & $1.027^{+0.003}_{-0.005}$\\
    $C_{\rm XRT}$ & $1.031\pm0.018$ & $1.031^{+0.017}_{-0.020}$ & $1.02\pm0.02$\\
    \hline
\red{$F_{\rm 1-78keV}$} & $2.193^{+0.008}_{-0.010}$ & $1.811^{+0.013}_{-0.012}$ & $1.866\pm0.013$ \\
    $\chi^{2}/\nu$ &  & 2907.72/2826 & \\
    \hline\hline
    \end{tabular}
\end{table}

So far, we have modelled the stacked spectra of \src\ in the hard state. By applying a lamppost model to the data, we obtain $R_{\rm in}<7$\,$r_{\rm g}$ and $i=12^{\circ}-29^{\circ}$. Consistent measurements are achieved when a power-law emissivity profile or a high density disc model is considered, suggesting the measurements are independent from the coronal geometry and the density parameter. 

In this section, we apply the \texttt{relxilllpcp} to each individual epoch to study their spectral variability. Only an upper limit of $n_{\rm e}$ is found for the stacked spectra. Therefore, we do not consider a variable density parameter in the following analysis. The inclination angle and the iron abundances are not expected be variable on observable timescales. Therefore, they are linked between epochs. The best-fit parameters of the \texttt{relxilllpcp} models for three epochs are shown in Table\,\ref{tab_3obs}. Best-fit models are shown in Figure\,\ref{pic_3obs}. The \texttt{relxilllpcp} model provides a good fit to all three sets of spectra. 

Based on our best-fit models, we find that the unabsorbed X-ray flux of \src\ varies between \red{$2.3-2.8\times10^{-9}$\,erg\,cm$^{-2}$\,s$^{-1}$ in the 0.01--100\,keV band}. Assuming a distance of 8\,kpc\footnote{Based on the stellar populations of the Galactic disc and bulge \citep{juric08}, a source along the line of sight of \src\ has a likely distance of $8^{+6}_{-2}$ kpc (1$\sigma$) excluding the effect of any natal supernova kicks \citep{russell18}.} \citep{russell18}, they correspond to \red{1.8--2.1$\times10^{37}$\,erg\,s$^{-1}$}. The mass of the BH in \src\ is unknown. Assuming a typical BH mass of 10\,$M_{\odot}$, \src\ is accreting at an Eddington ratio of \red{$\lambda_{\rm Edd}=1-2\%$}.

We find that most of the model parameters are consistent among three epochs, e.g. the inner radius of the disc, the height of the corona and the reflection fraction of the reflection component. 
The main difference of the spectra from the three observations is the photon index of the coronal Comptonisation continuum emission. The third observation has the softest continuum emission of $\Gamma\approx1.68$ while the first observation has the hardest continuum emission of of $\Gamma\approx1.62$.

\section{Discussion}

We analyse the \nustar\ and \swift\ spectra of \src\ in outburst in 2018. During the outburst, \src\ remains in the canonical hard state. The \nustar\ observations of \src\ show evidence of reflected emission from the inner region of the accretion disc. 

By modelling the reflection spectra with a lamppost model, we find a disc inner radius of $R_{\rm in}<7$\,$r_{\rm g}$ and a small inclination angle of around $23^{\circ}$. The abundances are found to be close to solar values. This suggests either a slightly truncated disc or a non-truncated disc forms at a few per cent of the Eddington limit in \src. The measurements of these parameters are consistent when a power-law emissivity profile or a high disc density is applied instead.

\subsection{The Disc Reflection Spectrum of \src}

In Section\,\ref{ne}, we consider a disc reflection model with a variable density parameter, although such a model does not significantly improve the fit. The data show tentative evidence of a high disc density of $n_{\rm e}\approx10^{19}$\,cm$^{-3}$ in \src. But only an upper limit of $n_{\rm e}\approx10^{21}$\,cm$^{-3}$ (90\% confidence range) is obtained. When a $n_{\rm e}\approx10^{19}$\,cm$^{-3}$ model is considered, the key parameters of the model, e.g. $i$ and $R_{\rm in}$, are consistent with those achieved by a low-density disc model.

The reflection models for the three epochs are mostly consistent, \second{suggesting} the same geometry of the innermost accretion for the period of \nustar\ observations. The inner radius of the disc remains a small value with an upper limit of 9--14\,$r_{\rm g}$, and the corona remains within a region of $\approx$10--20\,$r_{\rm g}$. The disc reflection fraction parameter is consistently around 0.18 for all three epochs. Therefore, we conclude that the multi-epoch variability of \src\ observed by \nustar\ is dominated only by the variable photon index of the X-ray continuum emission.

\subsection{The Inner Radius of the Disc in \src}

\red{Recent work on reflection modelling of BH X-ray binaries in the hard state focuses on the measurement of the inner disc radius. Whether the disc is significantly truncated or close to the ISCO at a modest luminosity, e.g. $L_{\rm X}=0.01-0.1 L_{\rm Edd}$, remains a disputed question.} 

\red{For example, \citet{garcia15} analysed the \textit{RXTE} spectra of GX~339$-$4 in the hard state. They found that the disc inner radius of this object moves outwards to 4.6\,$r_{\rm g}$ when its luminosity decreases from around 17\% to 1\% of Eddington. At the highest luminosity, the inner radius is consistent with the ISCO for a high BH spin of $a_{*}\approx0.95$. Similar results were found in \citet{reis08,wang17,jiang19} where different models and data were used. In particular, \citet{jiang19} fit the high density disc reflection model to the \nustar\ observations of this object in 2013 and 2015. In the 2013 outburst, GX~339$-$4 failed to transit to the soft state. No significantly different measurements of $R_{\rm in}$ were found in the full outburst and the failed-transition outburst of this object \citep[see Fig.\,8 in][]{jiang19}.}

\red{In comparison, \citet{plant15} found the disc in GX~339$-$4 is extremely truncated at 300\,$r_{\rm g}$ by fitting its \xmm\ observations in the hard state. A similar conclusion was found in \citet{kolehmainen14}. The disagreement between two completely different conclusions may be due to the calibration issues of the timing mode data with pile-up effects from \xmm\ \citep{garcia15}, which was noted by \citet{kolehmainen14}.}

\red{Similar efforts have been made for other objects too. For instance, GRS~1716$-$249 has been found to show a small inner radius of $<20r_{\rm g}$ by \citet{jiang20}. \citet{tao19}, however, found a much tighter constraint based on the same observation. They argued that the inner disc is consistent with ISCO for a high BH spin of >0.92. Different models were considered in these two pieces of work. The former modelled only the disc reflection spectra in GRS~1716$-$249. The latter was obtained by fitting both the disc thermal and reflection spectra. Instead of applying a Comptonised disc model to the disc thermal emission \citep[e.g.][]{steiner10}, \citet{tao19} fit the unscattered disc thermal emission component with a relativistic disc model. Disagreement was found in other objects too, e.g. MAXI~J1820$+070$ \citep{buisson19,zdziarski21} and XTE~J1752$-$223 \citep{garcia18,zdziarski21b}.}

\red{We report the first measurements of $R_{\rm in}$ in \src\ using reflection spectroscopy. The observations were taken during the hard state of this source when $L_{\rm X}\approx1-2\%L_{\rm Edd}$. We find all three epochs are consistent with an inner disc radius smaller than $9-15$\,$r_{\rm g}$. Assuming a maximum BH spin, the disc is either consistent with ISCO or slightly truncated. The choice of BH spin in our model does not affect our measurements of $R_{\rm in}$ (see Appendix\,\ref{a05}). By stacking the spectra of three epochs, we obtain a tighter constraint of $R_{\rm in}$ (<7\,$r_{\rm g}$). This result is similar to the measurements for some other objects at a similar Eddington ratio \citep[e.g.][]{wang12,xu18,jiang20}.}

\second{ Lastly, we note that previous analysis for some other sources in the hard state obtained a much tighter constraint on the disc inner radius and a more compact coronal geometry \citep[e.g.][]{fabian12,parker15,wang17,xu18}. Moreover, \citet{xu18} found that the broken power-law and lamppost emissivity profiles offer a different measurement for the disc inclination angle of MAXI~J1535$-$571. Differences due to systematic uncertainties in the disc reflection spectroscopy are expected when different flavours of reflection models are used \citep[e.g. see the latest review in ][]{bambi21}.}

\second{However, we can only achieve an upper limit of $R_{\rm in}<7$\,$r_{\rm g}$ for \src\  (see Table\,\ref{tab_cp}). Our model also suggests a slightly extended corona of $h\approx10-20$\,$r_{\rm g}$. Statistical uncertainties dominate over the systematic uncertainties of reflection spectroscopy in our case, because the contribution of the disc reflection component to the total X-ray flux is low in \src. For instance, $f_{\rm refl}$ is 1.5 (0.6) when a lamppost (broken power-law) emissivity profile is applied to the hard state observation of MAXI~J1535$-$571. In comparison, $f_{\rm refl}=0.19$ (0.08) is found for \src\ when a lamppost (broken power-law) emissivity profile is used. The low $f_{\rm refl}$ in \src\ might result from its slightly larger coronal region than the ones in other sources \citep[e.g. $h<10$\,$r_{\rm g}$,][]{fabian12,parker15,xu18}.} 

\section{Conclusions} \label{conclude}

\red{We have performed a detailed analysis of the \nustar\ and \swift\ spectra from observations of the hard state of \src. During the period of observations, the X-ray luminosity was 1--2\% of Eddington. The observed broad Fe~K emission cannot be modelled by narrow reflection features. A relativistic disc reflection model is thus applied to the data. We find, in agreement with several objects in a similar Eddington ratio \citep[e.g.][]{garcia15}, that the inner radius of the disc is close to the ISCO ($R_{\rm in}<7$\,$r_{\rm g}$, 90\% confidence uncertainty range). Multi-epoch spectral analysis is also performed. We find that the spectral difference between epochs results from the variable primary continuum emission from the corona while the geometry of the innermost accretion region remains consistent.}

\section*{Acknowledgements}

This paper was written during the worldwide COVID-19 pandemic in 2020--2022. We acknowledge the hard work of all the health care workers around the world. We would not be able to finish this paper without their protection. \red{J.J. acknowledges support from the Leverhulme Trust, the Isaac Newton Trust and St Edmund's College, University of Cambridge.} 

\section*{Data Availability}

All the data can be downloaded from the HEASARC website at https://heasarc.gsfc.nasa.gov. \second{The \texttt{reflionx} and \texttt{relxill} models used in this work are available for downloads at https://www.michaelparker.space/reflionx-models and http://www.sternwarte.uni-erlangen.de/dauser/research/relxill.}




\bibliographystyle{mnras}
\bibliography{iras13224.bib} 


\appendix

\section{Thermal emission in the hard state of \src} \label{bb_sec}

\begin{figure}
    \centering
    \includegraphics[width=8cm]{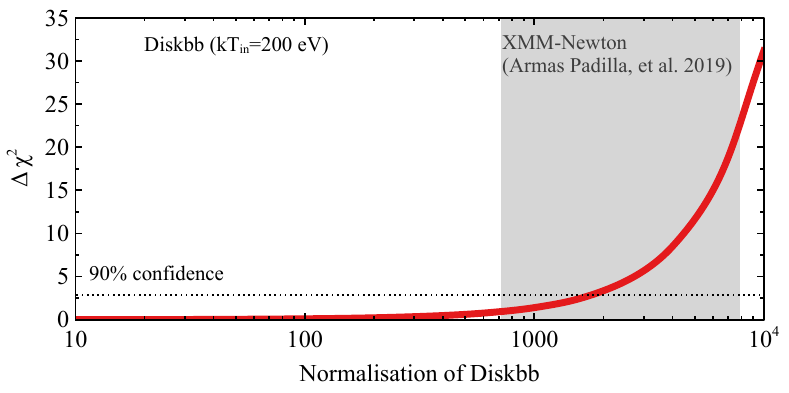}
    \caption{$\chi^{2}$ distribution against the normalisation parameter of the \texttt{diskbb} model for the stacked spectra of \src. The temperature is fixed at 200\,eV, the best-fit value for the \xmm\ observations of \src\ \citep{armas19}. Only an upper limit of 2000 is obtained for this component, corresponding to 1.8\% of the total X-ray luminosity.The grey shaded region shows the uncertainty range of the normalisation parameter given by the \xmm\ observations of the same source.}
    \label{pic_diskbb}
\end{figure}

Archival \xmm\ timing-mode observations of \src\ at the beginning of the outburst show some evidence of weak thermal emission in the soft X-ray band. This thermal component is consistent with disk blackbody emission of $kT_{\rm in}\approx200$\,eV \citep{armas19}. However, the contribution of this component to the total X-ray luminosity is low, e.g. <2\% \citep{armas19}.

We estimate the upper limit of such a thermal component in our data by adding an additional \texttt{diskbb} model. The $kT_{\rm in}$ parameter is fixed at 200\,eV, the value inferred from \xmm\ observations \citep{armas19}. We show the $\chi^{2}$ distribution against the normalisation parameter of \texttt{diskbb}. Only an upper limit of 2000 (90\% confidence) is obtained. At this upper limit, the \texttt{diskbb} component takes up only 1.8\% of the total luminosity in the 0.01--1000\,keV band. We, therefore, conclude that there is no significant evidence of thermal emission in our observations.

\section{The inner radius of the disc and BH spin} \label{a05}

\begin{table}
    \centering
    \caption{\red{Best-fit parameters for the stacked spectra of \src. In this fit, the BH spin parameter is fixed at 0.5. The corresponding radius of the ISCO is 0.43\,$r_{\rm g}$.}}
    \begin{tabular}{cc}
    \hline\hline
    Parameters  & Values\\
    \hline
    $N_{\rm H}$ ($10^{21}$\,cm$^{-2}$) & $8.10^{+0.02}_{-0.03}$ \\
    \hline
    h ($r_{\rm g}$) & $12^{+7}_{-5}$  \\
    $i$ (deg) &  $23^{+12}_{-10}$  \\
    $R_{\rm in}$ ($r_{\rm g}$) & <7  \\
    $\Gamma$ & $1.645\pm0.006$ \\
    $kT_{\rm e}$ (keV) & >180\\
    $Z_{\rm Fe}$ ($Z_{\odot}$) & $1.6^{+0.4}_{-0.5}$ \\
    $f_{\rm refl}$ &  $0.19^{+0.04}_{-0.03}$ \\
    {$\log(\xi$/erg\,cm\,s$^{-2})$} & $3.09\pm0.07$ \\ 
    Norm ($10^{-3}$) & $5.8\pm0.2$\\
    \hline
    $C_{\rm FPMB}$ & $1.025\pm0.003$ \\
    $C_{\rm XRT}$ & $1.03\pm0.03$ \\
    \hline
    $\chi^{2}/\nu$ & 1168.40/1106 \\
    \hline\hline
    \end{tabular} \label{tab_a05}
\end{table}

\red{The BH spin parameter $a_{*}$ is fixed at $0.998$ in Section\,\ref{stack} to investigate the possibility of $R_{\rm in}$ in the range of small values. In this appendix, we study whether the choice of $a_{*}$ affects our measurements of $R_{\rm in}$.}

\red{We consider the same reflection model \texttt{relxilllpcp} as in Section\,\ref{stack}. $a_{*}$ is fixed at 0.5, the corresponding $R_{\rm ISCO}$ of which is 4.32\,$r_{\rm g}$. The lower limit of $R_{\rm in}$ is thus at 4.32\,$r_{\rm g}$. We obtain a similarly good fit to the spectra of \src\ with $\chi^{2}/\nu=1168.40/1106$. $\chi^{2}$ is slightly higher than the value in Section\,\ref{stack} where $a_{*}=0.998$. Best-fit parameters are shown in Table\,\ref{tab_a05} and best-fit model is shown in Fig.\,\ref{tab_a05}.}

\begin{figure}
    \centering
    \includegraphics{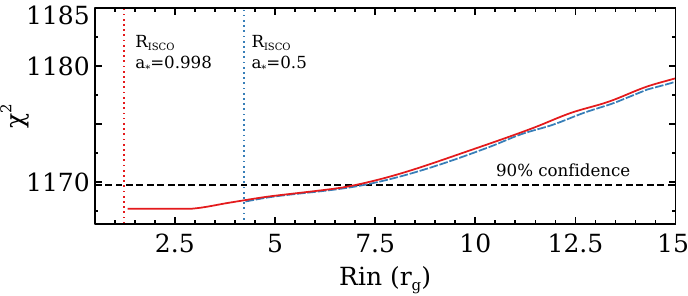}
    \caption{\red{$\chi^{2}$ as a function of $R_{\rm in}$ with different assumptions of $a_{*}$ in the reflection model (red solid line: $a_{*}=0.998$; blue dashed line: $a_{*}=0.5$). Two models provide similar constraints on $R_{\rm in}$. The red and blue vertical dotted lines show the radius of the ISCO for $a_{*}=0.998$ ($R_{\rm ISCO}=1.23r_{\rm g}$) and $a_{*}=0.5$ ($R_{\rm ISCO}=4.23r_{\rm g}$).}}
    \label{pic_rin_a05}
\end{figure}

\red{The $a_{*}=0.5$ and $a_{*}=0.998$ models provide consistent measurements of all the parameters. In particular, we show $\chi^{2}$ distribution of $R_{\rm in}$ in Fig.\,\ref{pic_rin_a05}. The 90\% confidence uncertainty ranges of $R_{\rm in}$ by two models are consistent. The difference is that the $a_{*}=0.998$ model allows $R_{\rm in}$ to be lower than 4.32\,$r_{\rm g}$. In conclusion, the measurement of $R_{\rm in}$ is unaffected by our choice of $a_{*}=0.998$ in Section\,\ref{stack}. }

\begin{figure}
    \centering
    \includegraphics{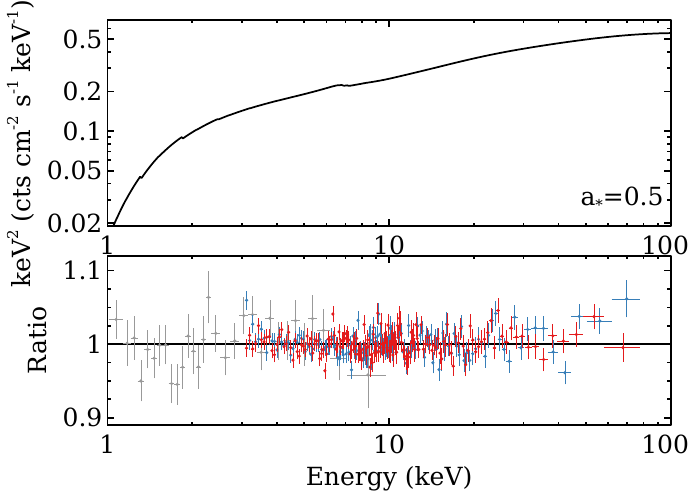}
    \caption{\red{Top: best-fit \texttt{relxilllpcp} model. Bottom: corresponding data/model ratio (red: FPMA; blue: FPMB; grey: XRT). In this model, we fix $a_{*}$ at 0.5. The \texttt{relxilllpcp} model with $a_{*}=0.998$ is used in Fig.\,\ref{pic_lpcp_fit}. Two models provide a similar fit to the spectra of \src.}}
    \label{pic_a05}
\end{figure}

\bsp	
\label{lastpage}
\end{document}